\documentclass[twocolumn,showpacs,amsmath,amssymb,aps,pra]{revtex4}
\newcommand{\half}{\mbox{$\frac{1}{2}$}}
\usepackage{amsmath}
\usepackage{amssymb}
\usepackage{graphics}
\usepackage{bm}
\begin{document}
\title{Stability, complex modes and non-separability in rotating quadratic
       potentials}
\author{R. Rossignoli, A.M. Kowalski}
\affiliation{Departamento de F\'{\i}sica-IFLP,
Universidad Nacional de La Plata, C.C.67, La Plata (1900),Argentina}
\begin{abstract}

We examine the dynamics of a particle in a general rotating quadratic
potential, not necessarily stable or isotropic, using a general complex mode
formalism. The problem is equivalent to that of a charged particle in a
quadratic potential in the presence of a uniform magnetic field. It is shown
that the unstable system exhibits a rich structure, with complex normal modes
as well as non-standard modes of evolution characterized by equations of motion
which cannot be decoupled (non-separable cases). It is also shown that in some
unstable cases the dynamics can be stabilized by increasing the magnetic field
or tuning the rotational frequency, giving rise to dynamical stability or
instability windows. The evolution in general non-diagonalizable cases is as
well discussed.
\end{abstract}
\maketitle

\section{Introduction}
Quadratic forms in boson operators or generalized coordinates and momenta  are
an ubiquitous presence in the theoretical description of diverse physical
systems. They often arise through the linearization of the equations of motion
around a stationary point, as in the case of the random phase approximation
(RPA)  \cite{RS.80,BR.86}, providing a basic tractable scenario. They  play an
important role in the description of Bose-Einstein condensates (BEC)
\cite{GM.97,PB.97,AK.02,F.07,E.07,F.08} as well as in other areas like quantum
optics \cite{Sa.91}, disordered systems \cite{GC.02} and dynamical systems
\cite{P.97, Do.00, Ko.00}.  Nonetheless, while positive forms, characteristic
of stable systems, are well known to be diagonalizable, i.e., they can be
written in terms of normal coordinates and viewed as a set of independent
bosons or separate oscillators (plus eventually free particle terms in the
presence of standard zero frequency modes \cite{RS.80,BR.86}), non-positive
ones may not admit such a diagonal representation \cite{RK.05}. Non-positive
forms can arise in the description of BEC instabilities \cite{F.07,E.07,F.08}
and fast rotating condensates \cite{D.05,BDS.08,Ft.01,Ft.07,O.04,A.09}, as well
as in generalized RPA treatments \cite{A.97,RC.97}.

In \cite{RK.05} we have extended the standard methodology for diagonalizing
quadratic bosonic forms by using generalized quasiparticle operators fulfilling
boson-like commutation relations, associated with non-hermitian coordinates and
momenta. This allows to characterize the operators exhibiting purely
exponential evolutions (complex modes) in non-positive forms, enabling a
precise description of the dynamics and quadratic invariants. We have also
pointed out that non-positive forms can  in some cases  be {\it dynamically
stable}, as the evolution can remain quasiperiodic, irrespective of the initial
conditions. Moreover, we have noticed the existence of non-diagonalizable cases
where the equations of motion cannot be fully decoupled and which may arise
even if all eigenfrequencies are non-zero. The method of ref.\ \cite{RK.05} has
been found useful in the context of BEC, being employed to study the emergence
of instabilities in trapped BEC with a highly quantized vortex
\cite{F.07,E.07,F.08} through the Bogoliubov-de Gennes equations.

Here we will apply this methodology to the basic problem of a particle in a
rotating anisotropic quadratic potential, {\it not necessarily stable}. This
system is formally equivalent to that of a charged particle in a uniform
magnetic field in a general quadratic potential \cite{FK.70,D.05,BDS.08}. The
problem is therefore relevant for many fields. In particular, the rotating case
has recently become relevant in the context of BEC in rotating anisotropic
traps \cite{D.05,BDS.08,Ft.01,Ft.07,O.04,A.09}, which in the Landau level
approach are basically described by a cranked quadratic potential of the type
here considered. The stable system is well known \cite{RS.80,BR.86,FK.70,V.56}
and diverse aspects of the stable anisotropic rotating case in the context of
rotating condensates have recently been investigated
\cite{Ft.01,Ft.07,O.04,A.09}.

We will here examine the general unstable case, which is of interest for fast
rotating condensates as the Hamiltonian ceases to be positive definite at high
frequency due to the centrifugal force. As we shall see, the unstable system
exhibits a rich structure, with several different dynamical regimes as well as
some quite remarkable features, including: a) The possibility of becoming
non-separable at the boundaries of regions with distinct dynamics, in the sense
that the Hamiltonian can no longer be written as a sum of two independent
standard or complex modes. In such cases the system will exhibit {\it anomalous
evolutions} characterized by a set of linear equations which cannot be
decoupled and which may lead to coordinates and/or momenta evolving with terms
$\propto te^{\lambda t}$ or even $\propto t^3$; b) The possibility of achieving
{\it dynamical stability} in some unstable cases by increasing the magnetic
field or tuning the rotational frequency. In particular, a stable anisotropic
rotating potential becomes dynamically unstable just in a finite frequency
window, recovering dynamical stability at high rotational frequency, whereas an
unstable saddle type potential can become dynamically stable in a certain
frequency window.

In sec.\ II we briefly revisit the main features of the formalism, discussing
the concept of separability in generalized coordinates and momenta and the
evolution for general non-separable cases. The application is discussed in III
while conclusions are drawn in IV.

\section{Formalism}
We consider a general quadratic hamiltonian
\begin{subequations}
\label{1}
\begin{eqnarray} h&=&\half\sum_{i,j}\,T_{ij}p_ip_j+V_{ij}q_iq_j+
U_{ij}(q_ip_j+p_jq_i)\label{1a}\\
&=&\half R^{t}{\cal H}_cR\,,\;\;\;
R=\left(\begin{array}{c}q\\p\end{array}\right)\,,
\;\;{\cal H}_c=\left(\begin{array}{cc}V&U\\U^t&T\end{array}\right)
\,,\label{1b}
\end{eqnarray}
\end{subequations}
where $t$ denotes transpose, $T$, $V$ are symmetric matrices and $p$, $q$ are
hermitian coordinates and momenta satisfying $[p_i,q_j]=-i\delta_{ij}$,
$[q_i,q_j]=[p_i,p_j]=0$, i.e.,
\begin{equation}
RR^{t}-(RR^{t})^{t}={\cal M}_c,
 \;\;{\cal M}_c=i\left(\begin{array}{cc}0&1\\-1&0\end{array}
\right)\,.\label{mc}\end{equation}

The ensuing Heisenberg equations of motion lead to a closed set of linear
equations which can be written as
\begin{eqnarray}i\frac{dR}{dt}&=&-[h,R]=\tilde{\cal H}_c R\,,\label{4a}\\
\tilde{\cal H}_c&=&{\cal M}_c{\cal H}=i\left(\begin{array}{cc}U^t&T\\
-V&-U\end{array}\right)\,,\label{4}
\end{eqnarray}
where $\tilde{\cal H}_c$ represents the RPA matrix in coordinate representation
\cite{RK.05}. It completely determines the system dynamics. Its eigenvalues
come in pairs of opposite sign and can be complex in unstable systems.
Moreover, $\tilde{\cal H}_c$ can also be {\it non-diagonalizable} (as in the
case of free particles $U=V=0$, $T_{ij}=t_i\delta_{ij}$, although other cases
can also arise, as discussed later). A positive definite ${\cal H}_c$
($R^t{\cal H}_cR>0$ $\forall$ real $R\neq 0$) ensures a diagonalizable
$\tilde{\cal H}_c$ together with a real spectrum \cite{RS.80,BR.86,RK.05}
(standard stable case) but the converse is not  true.

Under a general linear canonical transformation $R={\cal U}R'$, with the matrix
${\cal U}$ satisfying ${\cal U}{\cal M}_c{\cal U}^t={\cal M}_c$ in order to
preserve Eq.\ (\ref{mc}), we have ${\cal H}_c'={\cal U}^t{\cal H}_c{\cal U}$
but $\tilde{\cal H}_c'={\cal M}_c{\cal H}'_c={\cal U}^{-1} \tilde{\cal
H}_c{\cal U}$, ensuring the invariance of the eigenvalues and the Jordan
canonical form of $\tilde{\cal H}_c$.  Matrices $\tilde{O}\equiv {\cal M}_c O$
are precisely those accounting for the closed algebra of the forms (\ref{1}):
If $O_i=\half R^{\rm t}{\cal O}_iR$, then
\begin{equation}
[O_i,O_j]=\half R^{\rm t}{\cal C}R,\;\;{\rm with}\;\;
 \tilde{\cal C}=[\tilde{\cal O}_i,\tilde{\cal O}_j]\,.\label{C}\end{equation}

It is obviously equivalent to use a representation of $h$ in terms of boson
operators $b_j,b^\dagger_j=(q_j\pm ip_j)/\sqrt{2}$ satisfying
$[b_i,b^\dagger_j]=\delta_{ij}$. Defining $Z=(^{b}_{b^\dagger})$ and the
unitary matrix ${\cal S}=\frac{1}{\sqrt{2}}(^{1\;i}_{1\;-i})$, such that
$Z={\cal S}R$, we may rewrite $h$ as
\begin{eqnarray}
h&=&\sum_{i,j}A_{ij}(b^\dagger_ib_j+\half\delta_{ij})
+\half(B^+_{ij}b^\dagger_ib^\dagger_j+B^-_{ij}b_ib_j)\nonumber\\
&=&\half Z^\dagger {\cal H}Z,\;\;\;\;{\cal H}={\cal S} {\cal H}_c S^\dagger
=\left(\begin{array}{cc}A&B^+\\B^-&A^t\end{array}\right)\,,\label{bos}\\
A&=&\half[V+T-i(U-U^t)],\;\;B^{\pm}=\half[V-T\pm i(U+U^t)]\,.\nonumber
 \end{eqnarray}
The ensuing RPA matrix, defined by $idZ/dt=\tilde{\cal H}Z$, is just
$\tilde{\cal H}={\cal S} \tilde{\cal H}_c {\cal S}^\dagger$ and has obviously
the same eigenvalues (and Jordan canonical form) as $\tilde{\cal H}_c$.

\subsection{General evolution and dynamical stability}

For a time independent $h$, the solution of system (\ref{4a}) is
\begin{equation}R(t)=\exp[-i\tilde{H}_ct]R\,,\label{exp}\end{equation}
where $R\equiv R(0)$. Eq.\ (\ref{exp}) is itself a linear canonical
transformation. A system which is {\it dynamically stable}, i.e., leading to a
bounded quasiperiodic evolution of {\it all} operators $p_i$, $q_i$,
corresponds to a matrix $\tilde{\cal H}_c$  which i) {\it is diagonalizable}
and ii) {\it has only real eigenvalues}:

a) If $\tilde{\cal H}_c$ is diagonalizable, such that  $\tilde{\cal H}_c={\cal
W}\tilde{\cal H}'_c{\cal W}^{-1}$ with $(\tilde{\cal
H}'_c)_{\mu\nu}=\lambda_\nu\delta_{\mu\nu}$, we may expand Eq.\ (\ref{exp}) as
\begin{equation}
R(t)=\sum_\nu e^{-i\lambda_\nu t}{\cal W}_\nu Z'_\nu\,,\label{hdg}
\end{equation}
where ${\cal W}_\nu$ is the $\nu^{\rm th}$ column of the eigenvector matrix
${\cal W}$ and $Z'={\cal W}^{-1}R$ is a set of normal operators satisfying
\begin{equation}
i\frac{dZ'_\nu}{dt}=\lambda_\nu Z'_\nu\,,\label{Zp}
\end{equation}
and evolving then as $Z'_\nu(t)=e^{-i\lambda_\nu t}Z'_\nu$. For complex
eigenvalues $\lambda_\nu$, these generalized normal operators represent
exponentially increasing or decreasing modes (complex modes) and the dynamics
is unbounded ($\pm\lambda_\nu$ are both eigenvalues). They can always be
ordered and normalized such that those associated with $\pm \lambda_\nu$
($b'_\nu$ and $\bar{b}'_\nu$, with $Z'=(b',\bar{b'})^t$) satisfy boson-like
commutation relations \cite{RK.05}, i.e.,
$[b'_\nu,\bar{b}'_\mu]=\delta_{\mu\nu}$,
$[b'_\mu,b'_\nu]=[\bar{b'}_\mu,\bar{b'}_\nu]=0$, but $\bar{b'}_\nu\neq
{b'}_\nu^\dagger$ if $\lambda_\nu$ is non-real \cite{RK.05}.

b) If $\tilde{\cal H}_c$ is non-diagonalizable, the system (\ref{4a}) {\it
cannot be fully decoupled}, but we may use its Jordan canonical form
$\tilde{\cal H}_c={\cal W}\tilde{\cal H}'_c{\cal W}^{-1}$, with $\tilde{\cal
H}'_c$  having blocks of the form
\[\tilde{\cal H}'_c=\left(\begin{array}{ccccccc}
\ldots&&&&&&\\&\lambda_\nu&1&0&&0&\\
&0&\lambda_\nu&1&&0&\\&&&&\ddots&&\\&&&&&\lambda_\nu&\\
&&&&&&\ldots\end{array}\right)\,.\]
We may then expand Eq.\ (\ref{exp}) as
\begin{equation}
R(t)=\sum_{\nu}e^{-i\lambda_\nu t}\sum_{k=1}^{d_\nu}
{\cal W}_{\nu_k}\sum_{l=k}^{d_\nu}
Z'_{\nu_l} \frac{t^{l-k}}{(l-k)!}\label{jf}\,,
\end{equation}
where again $Z'={\cal W}^{-1}R$, $d_\nu$ is the dimension of the block and
$\nu_k$, $k=1,\ldots,d_\nu$ labels elements within each block, with ${\cal
W}_{\nu_k}$ the ${\nu_k}$ column of the generalized eigenvector matrix ${\cal
W}$. The generalized normal operators $Z'$ satisfy the ``minimally coupled''
evolution equations allowed by the Jordan form, i.e., $id Z'/dt=\tilde{\cal
H}'_c Z'$ or
\begin{equation}
i\frac{d Z'_{\nu_k}}{dt}=\lambda_\nu {Z}'_{\nu_k}
+(1-\delta_{k,d_\nu}){Z}'_{\nu_{k+1}}\,,
\end{equation}
leading to ${Z}'_{\nu_k}(t)=e^{-i\lambda_\nu t}
\sum_{l=k}^{d_\nu}{Z}'_{\nu_l}\frac{t^{l-k}}{(l-k)!}$. The dynamics is then
unbounded even for real $\lambda_\nu$. The free particle case corresponds to
$\lambda_\nu=0$ and $d_\nu=2$. Other cases are discussed in sec.\ III.

\subsection{Separability}
We will denote the Hamiltonian (\ref{1}) as {\it separable} if there is a
linear canonical transformation $R={\cal U}R'$ (with ${\cal U}{\cal M}_c{\cal
U}^t={\cal M}_c$) such that ${\cal H}'_c={\cal U}^t{\cal H}_c{\cal U}$ is {\it
diagonal}. In this case we may then rewrite $h$ as a sum of independent
elementary quadratic systems,
\begin{eqnarray} h&=&\half\sum_{\nu} (\alpha_\nu{p'}_\nu^2+
\beta_\nu {q'}_\nu^2)\,,\label{hd}
\end{eqnarray}
where $[q'_\nu,p'_{\mu}]=i\delta_{\nu\mu}$,
$[q'_\nu,q'_\mu]=[p'_\nu,p'_\mu]=0$. The diagonal form (\ref{hd}) is not
unique, as $p'_\nu$, $q'_\nu$ can be rescaled ($(p'_\nu,q'_\nu)\rightarrow
(\alpha p'_\nu,q'_\nu/\alpha)$) or swapped ($(p'_\nu,q'_\nu)\rightarrow
(q'_\nu,-p'_\nu)$),  but the products $\alpha_\nu\beta_\nu=\lambda_\nu^2$
determine the eigenvalues of $\tilde{\cal H}_c$ and are hence unique.

In contrast with the conventional normal mode expansion of a positive definite
$h$, each of the terms in (\ref{hd}) can here represent not only {\bf i)} a
standard stable oscillator ($\alpha_\nu>0$, $\beta_\nu>0$), but also {\bf ii)}
an ``inverted'' oscillator ($\alpha_\nu<0$, $\beta_\nu<0$), {\bf iii)} a
generalized free particle ($\alpha_\nu\beta_\nu=0$, with $\alpha_\nu\neq 0$ or
$\beta_\nu\neq 0$, which can be standard or inverted), {\bf iv)} an ``unstable
oscillator'' ($\alpha_\nu\beta_\nu<0$) and {\bf v)} a ``complex oscillator''
($\alpha_\nu\beta_\nu$ complex), where $p'_\nu$, $q'_\nu$ {\it are no longer
hermitian} (${\cal U}$ complex). We should also add the {\it vanishing} case
{(0)} $\alpha_\nu=\beta_\nu=0$, where {\it both} $p'_\nu$ and $q'_\nu$
commute with $h$. Separability in hermitian coordinates and momenta (${\cal U}$
real) is  a {\it restricted} class of separability \cite{RK.05}, since the
eigenvalues $\pm \sqrt{\alpha_\nu\beta_\nu}$ of $\tilde{\cal H}_c$ are in such
a case real or imaginary, while diagonalizable cases with full complex
$\lambda_\nu$ do exist \cite{RK.05}.

A diagonalizable $\tilde{\cal H}_c$ ensures separability since in this case we
may rewrite $h$ in terms of the generalized normal operators
$Z'=(b',\bar{b}')^t$ of Eq.\ (\ref{Zp}) as \cite{RK.05}
\begin{eqnarray}
h&=& \sum_\nu\lambda_\nu(\bar{b}'_\nu b'_\nu+\half) =\half
 \sum_\nu\lambda_\nu({p'}_\nu^2+{q'}_\nu^2)\,, \label{hdo}\end{eqnarray}
where $q'_\nu=(b'_\nu+\bar{b'}_\nu)/\sqrt{2}$,
$p'_\nu=(b'_\nu-\bar{b'}_\nu)/(\sqrt{2}i)$, i.e., ${\cal U}={\cal W}{\cal S}$. For
complex $\lambda_\nu$, $\bar{b}'_\nu\neq {b'}_\nu^\dagger$ and $p'_\nu,q'_\nu$
are non-hermitian. In case iv) we may still rewrite the ensuing term in
(\ref{hdo}) in terms of hermitian $q'_\nu$, $p'_\nu$ by a complex rescaling
$(p'_\nu,q'_\nu)\rightarrow (p'_\nu/\sqrt{i},\sqrt{i}q'_\nu)$ \cite{RK.05},
while in ii) we should choose $\lambda_\nu<0$ for $p'_\nu$, $q'_\nu$ hermitian.
Conversely, for $\alpha_\nu\beta_\nu\neq 0$ or $\alpha_\nu=\beta_\nu=0$, each
term in (\ref{hd}) leads to a diagonalizable $2\times 2$ block in $\tilde{\cal
H}'_c$. However, the separable case also includes {\it the free particle case}
iii) where $\tilde{\cal H}_c$ is {\it non-diagonalizable}, as  $\tilde{\cal
H}'_c$ will contain a Jordan block $\tilde{\cal H}'_0=(^{0\;1}_{0\;0})$ with
$\lambda_\nu=0$. Here $\tilde{\cal H}_c^2$ remains diagonalizable.

Hence, systems where $\tilde{\cal H}_c^2$ is {\it non-diagonalizable}, implying
$\tilde{\cal H}'_c$ having a Jordan block of dimension $d_\nu>2$ or $d_\nu=2$
and $\lambda_\nu\neq 0$, are {\it non-separable}. They may arise even in simple
unstable cases (sec.\ III) and their evolution can be determined through the
general solution (\ref{jf}).

Dynamically stable quadratic systems correspond  to a separable $h$ with terms
just of the form  i), ii) or 0), and have then a discrete spectrum
\begin{equation}
E_{\{n_\nu\!\}}={\textstyle\sum_\nu} \lambda_\nu(n_\nu+\half)\,,\label{en}
\end{equation}
where $\lambda_\nu>0$, $<0$ or 0 in cases i), ii) or 0). An example of a
non-positive  dynamically stable form is an angular momentum component
$l=q_+p_--q_-p_+=\half({p'}_+^2+{q'}_+^2)-\half({p'}_-^2+{q'}_-^2)$, where
$p'_{\pm}=p_{\pm}-q_{\mp}/2$, $q'_{\pm}=q_\pm/2+p'_{\mp}$, which is the sum of
a positive plus an inverted oscillator. Here $\tilde{\cal L}_c$ is
diagonalizable with eigenvalues $\pm 1$, $\mp 1$.

\section{Application}
We will consider the quantum problem of a particle in an anisotropic quadratic
potential, {\it not necessarily stable}, rotating around one of its principal
axes ($z$). It is formally equivalent to that of a particle of charge $e$ in a
uniform magnetic field $\bm{H}$ parallel to this axis in a quadratic potential.
The Hamiltonian of the latter reads
\begin{eqnarray}
 H&=&\frac{(\bm{P}-e\bm{A}/c)^2}{2m}+\half(K_x X^2+K_y Y^2+K_z Z^2)
 \label{H1}\\
&=&\half\,[\frac{P_x^2+P_y^2}{m} +K'_x X^2+K'_y Y^2-\Omega L_z]+H_z\,,
\label{H2}
\end{eqnarray}
where $\bm{A}=\half\bm{H}\times\bm{R}$ is the vector potential, $\Omega=e
|\bm{H}|/mc$ the cyclotron frequency, $L_z=XP_y-YP_x$ the angular momentum
component, $H_z=\half(\frac{P_z^2}{m}+K_z Z^2)$ and
\[K'_{x,y}=K_{x,y}+m\Omega^2/4\,.\]
For $\Omega\rightarrow 2\Omega$, Eq.\ (\ref{H2}) is just the cranked
Hamiltonian describing the intrinsic motion of a particle in a rotating
quadratic potential with constants $K'_{x,y}$: If $H(t)=U(t) H(0)U^\dagger(t)$,
with $U(t)=e^{-i\Omega L_z t/\hbar}$, the Heisenberg equations for rotating
operators $O(t)=U(t)OU^\dagger(t)$ are those for the $t$-independent cranked
Hamiltonian $H=H(0)-\Omega L_z$.

Since $H_z$ is fully decoupled from the rest and its treatment is trivial, it
will  be omitted in what follows and all considerations will refer to the
motion in the $x,y$ plane. Defining dimensionless operators
$\bm{q}=\bm{R}\sqrt{m\Omega_0/\hbar}$, $\bm{p}=\bm{P}/\sqrt{\hbar m\Omega_0}$
satisfying $[q_\mu,p_{\mu'}]=i\delta_{\mu\mu'}$, where $\Omega_0$ is a
reference frequency, we can rewrite $h\equiv H_{xy}/\hbar\Omega_0$ as
\begin{eqnarray}
\label{hadim} h&=&\half[p_x^2 +p_y^2+k'_xq_x^2+k'_yq_y^2] -\omega
l_z\label{hm}\\&=&\half R^t {\cal H}_c R\,,\;\;
k'_{x,y}=k_{x,y}+\omega^2,\label{hm2}
\end{eqnarray}
with $k_{\mu}=K_{\mu}/(m\Omega_0^2)$,
$\omega=\Omega/(2\Omega_0)$, $l_z=q_x p_y - q_y p_x$ and 
$R^t=(q_x,q_y,p_x,p_y)$.

From the form of Eq.\ (\ref{H1}), it is apparent that for fixed $k_\mu$, the
field cannot change the number of positive or negative eigenvalues of the
Hamiltonian matrix ${\cal H}_c$ (the number of positive and negative diagonal
elements is the same in any real diagonal representation of a quadratic form).
For $m>0$, ${\cal H}_c$ will then have none,  one or two negative eigenvalues
if and only if none, one or both of the constants $k_x,k_y$ are, respectively,
negative. The positive definite case corresponds then to $k_x>0$, $k_y>0$
$\forall$ $\omega$, i.e., $k'_\mu>\omega^2$ \cite{Ft.01,Ft.07}, although we
will now see that dynamical stability {\it is not restricted to this case}.

\subsection{Dynamical stability}
\begin{figure} \vspace*{-2.75cm}

\centerline{\hspace*{.5cm}\scalebox{.55}{\includegraphics{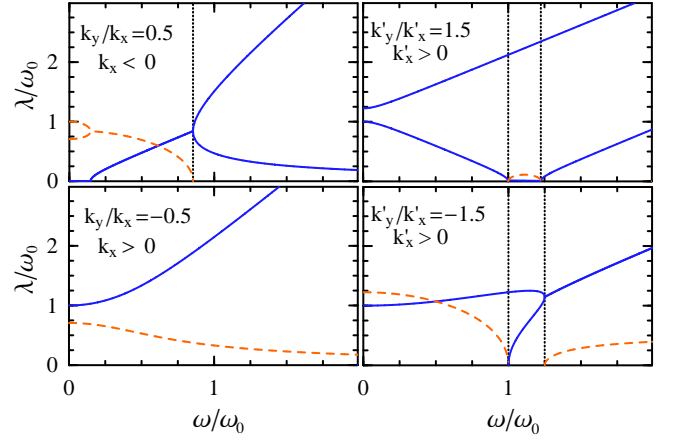}}}
\vspace*{-7.95cm}

\caption{The real (solid, blue lines) and imaginary (dashed, red
lines) parts of the eigenfrequencies (\ref{la}) for selected fixed values of
$k_{x,y}$ (left panels, corresponding to a particle in a magnetic field) and
$k'_{x,y}$ (right panels, corresponding to a particle in a rotating potential),
in terms of the scaled cyclotron or rotational frequency. Vertical dotted lines
separate dynamically stable and unstable regions. We have set
$\omega_0=\sqrt{|k_x|}$ ($\sqrt{k'_x}$) in the left (right) panels. }
\label{f1}\vspace*{-0.25cm}
\end{figure}

The RPA matrix $\tilde{\cal H}_c={\cal M}_c{\cal H}$ becomes
\begin{eqnarray}
\tilde{\cal H}_c&=&
i\left(\begin{array}{cccc}0 &\omega&1&0\\
-\omega&0&0&1\\-{k'}_x&0&0&\omega\\0&-{k'}_y&-\omega&0\end{array}\right)\,,
\label{RPA}\end{eqnarray}
and its eigenvalues, which are the system eigenfrequencies
(in units of $\hbar\Omega_0$), are $\pm\lambda_+,\pm\lambda_-$ with
\begin{eqnarray}\lambda_{\pm}&=&\sqrt{(k'_x+k'_y)/2
+\omega^2\pm\Delta}\,,\label{la}\\
\Delta&=&\sqrt{(k'_x-k'_y)^2/4+2\omega^2(k'_x+k'_y)}\,.\label{delta}
\end{eqnarray}
They satisfy $\lambda_+^2\lambda_-^2={\rm det}[{\cal H}_c]=k_x k_y$. We first
note that for $k_xk_y\neq 0$ and $\Delta\neq 0$, $\tilde{\cal H}_c$ {\it is
diagonalizable} since it will have four different eigenvalues. We will then
show that for $\Delta=0$, $\tilde{\cal H}_c$ is {\it non-diagonalizable} and,
moreover, $h$ is {\it non-separable}. At fixed $k_{\mu}$,  $\Delta=0$ if
$k_{x,y}<0$ and
\begin{equation} |\omega|=\omega_c^\pm=\half|\sqrt{-k_x}\pm\sqrt{-k_y}|
\,.\label{wc}
\end{equation}
It also follows that $\lambda_{\pm}$ can be {\it both} real only if $k_x
k_y\geq 0$. For a charged particle in a magnetic field,  this opens up the
possibility of full dynamical stability around a quadratic maximum
($k_{x,y}<0$) but dismisses it for a saddle point ($k_x k_y<0)$. It is indeed
verified that $\lambda_{\pm}$ are both real for $k_{x,y}>0$ {\it as well as for
$k_{x,y}<0$ and}
\begin{equation} |\omega|\geq\omega_c^+\,.\end{equation}
At fixed $k_\mu$, the dynamics in the vicinity of a quadratic maximum {\it can
then be stabilized by increasing the field}. The behavior of $\lambda_\pm$ for
increasing $|\omega|$ at fixed $k_{\mu}$ is depicted on the left panels of
Fig.\ \ref{f1}. As the scaled cyclotron frequency $2|\omega|$ is increased, at
an anisotropic maximum $\lambda_\pm$ evolve from imaginary
($|\omega|\leq\omega_c^-$) to full complex ($\omega_c^-<|\omega|<\omega_c^+$)
and finally to {\it real} ($|\omega|\geq\omega_c^+$) values, reaching  the
system dynamical stability for $|\omega|>\omega_c^+$, whereas for $k_xk_y<0$,
$\lambda_+$ is real but $\lambda_-$ is imaginary $\forall\omega$.

{\it Stability in rotating potential}. At {\it fixed} $k'_\mu$, the previous
picture is seen quite differently and leads to {\it dynamical stability and
instability windows} in the  anisotropic case $k'_x\neq k'_y$ (right panels in
Fig.\ \ref{f1}), i.e., when the rotation has a non-trivial effect. Owing to the
centrifugal force, ${\cal H}_c$ is here positive definite just for
$\omega^2<k'_{x,y}$. However, $\lambda_\pm$ are real {\it also for}
$\omega^2>k'_{x,y}>0$, implying that motion in a rotating stable potential {\it
becomes dynamically stable at high rotational frequencies}, and dynamically
unstable just in the {\it finite} interval
 \begin{equation} {\rm
Min}[k'_x,k'_y]\leq\omega^2\leq{\rm Max}[k'_x,k'_y]\,,\label{uns}
 \end{equation}
where $\lambda_-$ becomes imaginary or 0 (see III B).  In contrast, for a
saddle point with $k'_x>0>k'_y$ (or viceversa) the system {\it becomes
dynamically stable in the windows}
\begin{equation}\begin{array}{lcl}
k'_x<\omega^2&\;&(-k'_x<k'_y<0)\,,\\
k'_x<\omega^2<{\omega'_c}^2&\;&(-3k'_x<k'_y<-k'_x)\,,\end{array}
\label{w4}\end{equation}
with $\omega'_c=\frac{|k'_x-k'_y|}{\sqrt{-8(k'_x+k'_y)}}$ the single value of
$|\omega|$ where $\Delta=0$ at fixed $k'_{\mu}$ (requires $k'_x+k'_y<0$). Thus,
$l_z$ can turn unstable the dynamics of a stable anisotropic oscillator and
stabilize that around a saddle point within the limits (\ref{w4}).
It can never stabilize a quadratic maximum.

In the isotropic case $k'_x=k'_y=k'$, $[h,l_z]=0$ and Eq.\ (\ref{la}) leads
then to $|\lambda_{\pm}|=|\sqrt{k'}\pm\omega|$. The rotation has  no effect
except for the shift $\pm\omega$. The instability window collapses into a
single point $\omega^2=k'$ where the dynamics remains {\it stable} (see below),
since it  corresponds to the Landau case $k_x=k_y=0$ \cite{L.30,FK.70}.

\begin{figure}
\vspace*{-3.75cm}

\centerline{\hspace*{.5cm}\scalebox{.75}{\includegraphics{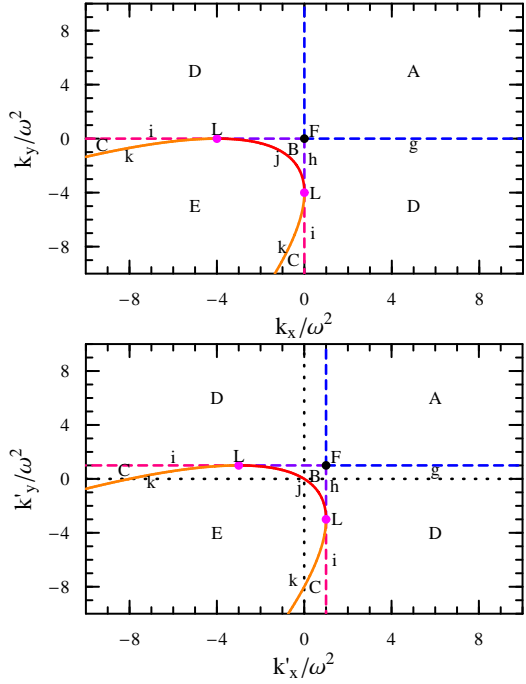}}}
\vspace*{-9.5cm}
\caption{Top: Regions in the $k_{x,y}$ plane at
fixed $\omega$ with distinct dynamical regimes: A is the positive definite
sector, F the Landau point, B the non-positive dynamically stable sector, C and
D are unstable regions with four and two imaginary eigenfrequecies, $E$ that
with full complex eigenfrequencies, while the solid lines j, k depict the {\it
non-separability curve} $\Delta=0$, with degenerate real (j) or imaginary (k)
eigenfrequencies (see text). The points $L$ ($k_x=0$, $k_y=-4\omega^2$ or
vice-versa) are the exceptional non-separable cases with a single vanishing
eigenfrequency. Dashed lines indicate separable cases with a free particle term
(standard in g,i and inverted in h) and a stable (g,h) or unstable (i)
oscillator. Bottom: Same details in the  $k'_{x,y}$ plane, corresponding to the
rotating system: All regions and curves are just shifted by $\omega^2$. Plots
in Fig.\ \ref{f1} depict the behavior of $\lambda_\pm$ along straight lines
running from $\infty$ ($\omega=0$) to the origin
($|\omega|\rightarrow\infty$).} \label{f2}\vspace*{-0.25cm}
\end{figure}

\subsection{Separability \label{S}}
Let us now examine the separable representation of Eq.\ (\ref{hm}),  feasible
for $\Delta\neq 0$, and the ensuing distinct dynamical regimes. Defining, as in
the stable case \cite{RS.80,BR.86},
\begin{eqnarray}
p_\pm&=&p_{x,y}+\gamma q_{y,x},\;\;q_\pm= \frac{q_{x,y}-\eta
p_{y,x}}{1+\gamma\eta}\,, \label{trnsf}\end{eqnarray} where
$\gamma=\frac{2\Delta-k'_x+k'_y}{4\omega}$, $\eta=\frac{2\gamma}{k'_x+k'_y}$
and $[q_\nu,p_\mu]=i\delta_{\mu\nu}$, $[p_\mu,p_\nu]=[q_\mu,q_\nu]=0$ for
$\mu,\nu=\pm$ (with $\gamma,\eta\rightarrow 0$ if $\omega\rightarrow 0$ and
$k'_x>k'_y$), we may rewrite (\ref{hm}) for $\Delta\neq 0$ as a sum of two
independent elementary quadratic forms ,
\begin{eqnarray}
h&=&\half(\alpha_+{p}_+^2+\beta_+{q}_+^2)
+\half (\alpha_-{p}_-^2+\beta_-{q}_-^2)\,,\label{hdu}\\
\alpha_\pm&=&{\textstyle\frac{2\Delta+k'_x-k'_y\pm 4\omega^2}{4\Delta},\;\;
\beta_\pm=\frac{\Delta(2\Delta-k'_x+k'_y\pm
4\omega^2)}{4\omega^2}}\,,\label{trnsf2}
\end{eqnarray}
with $\alpha_\pm\beta_\pm=\lambda_\pm^2$, For real $\Delta\neq 0$, $p_\pm$,
$q_\pm$ are hermitian  and $\lambda_{\pm}$ is real or imaginary. Eqs.\
(\ref{hdu}--\ref{trnsf2}) are, however, also applicable for {\it imaginary}
$\Delta\neq 0$, where $p_\pm, q_\pm$ {\it are non-hermitian} and $\lambda_\pm$
complex. From the previously mentioned property of diagonal quadratic forms, it
follows that all four coefficients $\alpha_\pm$, $\beta_{\pm}$ must be positive
for $k_{x,y}>0$,  just one (two) of them will be negative for $k_xk_y<0$
($k_{x,y}<0$ and $\Delta$ real), one of them will vanish for $k_y=0$, $k_x\neq
0$ (or vice-versa) whereas two of them will vanish in the Landau case
$k_x=k_y=0$.

{\it Diagonalizable cases}. For $\Delta\neq 0$ and $k_xk_y\neq 0$,
$\lambda_{\pm}\neq 0$ and $\tilde{\cal H}_c$ is diagonalizable. Defining
${p'}_\pm= \sqrt{\alpha_\pm/\lambda_\pm}\,{p}_\pm$,
${q'}_\pm=\sqrt{\beta_\pm/\lambda_\pm}\,{q}_\pm$,  we may rewrite (\ref{hdu})
as
\begin{eqnarray}
h&=&\half\sum_{\nu=\pm}{\lambda}_\nu({p'}_\nu^2+{q'}_\nu^2)
=\sum_{\nu=\pm} \lambda_\nu(\bar{b}'_\nu b'_\nu+\half)\,,
\label{nc1}
\end{eqnarray}
where
$b'_\nu=\frac{{q'}_\nu+i{p'}_\nu}{\sqrt{2}}$,
$\bar{b}'_\nu=\frac{{q'}_\nu-i{p'}_\nu}{\sqrt{2}}$ are the
generalized normal operators evolving as
$b'_\nu(t)=e^{-i\lambda_\nu
t}b'_\nu$, $\bar{b}'_\nu(t)= e^{i\lambda_\nu t}\bar{b}'_\nu$,
which can be directly obtained from the eigenvectors of $\tilde{\cal H}_c$.
At fixed $k_\mu$, the diagonalizable sectors are (fig.\ \ref{f2}):  \\
A) $k_{x,y}>0$: Here $\alpha_\pm>0$, $\beta_\pm>0$, with
$\lambda_{\pm}>0$. This is the positive definite case (case i in sec.\ II B).\\
B) $k_{x,y}<0$, $|\omega|>\omega_c^+$:  Here $\alpha_+>0$, $\beta_+>0$ but
$\alpha_-<0$, $\beta_-<0$, implying $\lambda_+>0$ but  $\lambda_-<0$ for
${p'}_-,{q'}_-$ hermitian. Eq.\ (\ref{nc1}) becomes a standard plus an inverted
oscillator (cases i+ii), remaining
{\it dynamically stable}.\\
C) $k_{x,y}<0$, $|\omega|<\omega_c^-$: Here $\alpha_\nu \beta_\nu<0$ for
$\nu=\pm$ and $\lambda_\pm$ are both imaginary. Both terms in (\ref{hdu}) are
unstable oscillators (case iv), leading to  ${p'}_{\pm}$,
${q'}_{\pm}$ non-hermitian. \\
D) $k_xk_y<0$: Here $\alpha_+>0,\beta_+>0$ but $\alpha_-\beta_-<0$, with
$\lambda_+$ real, $\lambda_-$ imaginary.  Eq.\ (\ref{hd}) becomes a stable plus
an unstable oscillator (i+iv), with
${p'}_-, {q'}_-$ non-hermitian. \\
In all previous cases $p_\pm$, $q_\pm$ in (\ref{hdu}) are hermitian.\\
E) $k_{x,y}<0$, $\omega_c^-<|\omega|<\omega_c^+$: Here $\alpha_\pm$,
$\beta_\pm$ and $\lambda_\pm$ are full complex and ${p}_\pm$, ${q}_{\pm}$ as
well as ${p'}_{\pm}$, ${q'}_{\pm}$ are non-hermitian (case v). They represent
complex normal modes.
\\
F) $k_x=k_y=0$ ({\it Landau case}): Here $\alpha_+=1$, $\beta_+=4\omega^2$
whereas $\alpha_-=\beta_-=0$, leading to $\lambda_+=2|\omega|$ and
$\lambda_-=0$. $h$ is then a standard plus a {\it vanishing} oscillator (cases
i + 0). This well known case \cite{L.30,FK.70} is then dynamically stable in
the $x,y$ plane despite the vanishing egenfrequency.

In cases A, B and F, ${p'}_\nu$, ${q'}_\nu$ are  hermitian, with
$\bar{b}'_\nu={b'}_\nu^\dagger$, and $h$ possesses then a discrete spectrum
\begin{equation} E_{n_+,n_-}={\lambda}_+n_++
{\lambda}_-n_-+\half({\lambda}_++{\lambda}_-)\end{equation} with $\lambda_+>0$,
while  ${\lambda}_->0$ in A, $\lambda_-<0$ in B and $\lambda_-=0$ in F. These
are the dynamically stable cases.

At fixed $k'_\mu$ (rotating potential) all regions are just shifted by
$+\omega^2$ (lower panel in fig.\ \ref{f2}). This shift leads to the different
behavior of $\lambda_\pm$ with $\omega$ depicted in fig.\ \ref{f1}.

{\it Separable non-diagonalizable cases}. They arise for $k_y=0$ and $k_x\neq
0$ or viceversa (sectors g,h,i in Fig.\ \ref{f2}). For $k_y=0$,
$\Delta=2\omega^2+k_x/2$ and we obtain $\lambda_-=0$ but
$\lambda_+=\sqrt{4\omega^2+k_x}\neq 0$ for $\Delta\neq 0$, with
\begin{equation}
h=\half({p}_+^2+\lambda_+^2{q}^2_+)+\half\frac{k_x}{\lambda_+^2}{p}_-^2
\label{hs0}\,.
\end{equation}
In g, $k_x>0$ and Eq.\ (\ref{hs0}) corresponds to a stable oscillator
($\lambda_+>0$) plus a free particle (cases i+iii). In h,  $-4\omega^2<k_x<0$
and $\lambda_+$ is still real, but the second term in (\ref{hs0}) becomes
negative: $h$ becomes  {\it a stable oscillator} plus an ``inverted'' free
particle term.  The latter ``absorbs'' here the instability,  {\it allowing
dynamical stability in the coordinates} $p_+,q_+$. Finally, in sector i
($k_x<-4\omega^2$)  $\lambda_+$ becomes imaginary: Eq.\ (\ref{hs0}) corresponds
here to  an unstable oscillator plus a standard free particle (iv+iii).

Eq.\ (\ref{hs0}) leads to an evolution
$\bar{b}'_+(t)=e^{i\lambda_+t}\bar{b}'_+, b'_+(t)=e^{-i\lambda_+t}b'_+$,
$p_-(t)=p_-$, $q_-(t)=(k_x/\lambda_+^2)p_-t+q_-$, which is characteristic of a
non-diagonalizable $\tilde{\cal H}_c$ with a canonical form (${\cal A}\sim
{\cal B}$ denotes ${\cal A}={\cal W}^{-1}{\cal B}{\cal W}$)
\begin{equation}\tilde{\cal H}_c\sim\left(\begin{array}{cccc}\lambda_+&0&0&0
\\0&-\lambda_+&0&0\\0&0&0&1\\0&0&0&0\end{array}\right)\label{sim0}\,.
\end{equation}

\subsection{Non-separability}
We now examine the very peculiar case $\Delta=0$ and $\omega\neq 0$, where the
eigenfrequencies become degenerate,
$\lambda_{\pm}=\sqrt{\frac{(3k'_x+k'_y)(3k'_y+k'_x)}{8(k'_x+k'_y)}}=\lambda$,
and  both $\tilde{\cal H}_c$ and  $\tilde{\cal H}_c^2$ are {\it
non-diagonalizable}. It occurs at the threshold for full complex solutions,
i.e., $|\omega|=\omega_c^\pm$ ($\omega'_c$) at fixed $k_\mu$ ($k'_\mu$) and
corresponds at fixed $\omega$ to a parabola rotated $\pi/4$ with respect to the
$k_{x,y}$ axes, with vertex at $k_{x,y}=-\omega^2$, i.e., $k'_{x,y}=0$ (curves
j, k, and point L in fig.\ \ref{f2}).

j) Here $|\omega|=\omega_c^+$ and  $\lambda=\sqrt[4]{k_xk_y}=
\sqrt{{\omega_c^+}^2-{\omega_c^-}^2}>0$. $\tilde{\cal H}_c$ can  be reduced to
two non-trivial Jordan blocks
\begin{equation}\tilde{\cal H}_c\sim\left(\begin{array}{cccc}
\lambda&1&0&0\\0&\lambda&0&0\\0&0&-\lambda&1\\0&0&0&-\lambda
  \end{array}\right)\label{sim1}\,,\end{equation}
which indicates non-separability.
The transformation
\begin{eqnarray}p_\pm&=&\frac{\sqrt{\omega_c^+}}{\sqrt[4]{-k_{x,y}}}
(p_{x,y}+\omega_c^- q_{y,x})\label{tt1}\\
q_\pm&=&\frac{\sqrt[4]{-k_{x,y}}}{\sqrt{\omega_c^+}}\frac{{\omega_c^+}^2
+\lambda^2}{2\lambda^2}(q_{x,y}+\frac{\omega_c^-}
{{\omega_c^+}^2+\lambda^2}p_{y,x})\label{tt2}
\end{eqnarray}
allows to express $h$ in this case as
\begin{equation}
h=\half({p}_+^2+{p}_-^2)-\lambda(q_+p_--q_-p_+)\,,\label{hs1}
\end{equation}
which is the cranked hamiltonian for a rotating free particle and is a basic
non-separable form: The equations of motion $dp_\pm/dt=\pm\lambda p_{\mp}$,
$dq_{\pm}/dt=\pm \lambda q_{\mp}+p_{\pm}$, cannot be fully decoupled even
though there is no vanishing eigenfrequency. More explicitly, defining the
operators ${b}'_{\pm}=\frac{\pm p_{+}+i p_-}{\sqrt{2}}$,
$\bar{b}'_{\pm}=\frac{\pm iq_{+}+q_-}{\sqrt{2}}$ satisfying
$[b'_\nu,\bar{b}'_{\mu}]=\delta_{\mu\nu}$,
$[\bar{b}'_\nu,\bar{b}'_{\mu}]=[b'_\nu,b'_{\mu}]=0$, we can rewrite Eq.\
(\ref{hs1}) as
\begin{equation}h=\lambda(\bar{b}'_+b'_+-\bar{b}'_-b'_-)
-b'_+b'_-\,.\label{hs12}
\end{equation}
The ensuing equations of motion, $i\frac{d{b}'_{\pm}}{dt}=\pm\lambda
{b}'_{\pm}$, $i\frac{d\bar{b}'_{\pm}}{dt}=\mp \lambda
\bar{b}'_{\pm}+{b}'_{\mp}$, correspond  exactly to the Jordan form (\ref{sim1})
and possess the general solution
\[{b}_{\pm}'(t)=e^{\mp i\lambda t}{b}'_{\pm},\;\;\bar{b}'_{\pm}(t)=
 e^{\pm i\lambda t}(\bar{b}'_{\pm}-i t{b}'_{\mp})\]
which gives rise to unbounded spiral-like trajectories in the variables
$p_\pm$, $q_\pm$. This is apparent from Eq.\ (\ref{hs1}): Since $[l,h]=0$,
where $l=q_+p_--q_-p_+$, the evolution operator $e^{-iht}=e^{i\lambda
lt}e^{-i(h+\lambda l) t}$ represents a rotation of frequency $\lambda$ applied
to a free motion. Such motion cannot arise from a separable $h$: Although $h$
is the sum of two commuting quadratic forms and the equations of motion can be
reduced to two separate blocks, the operators in each block do not commute with
those of the other.

k) Here $|\omega|=\omega_c^-$ and
$\lambda=i\sqrt[4]{k_xk_y}=\sqrt{{\omega_c^-}^2-{\omega_c^+}^2}\neq 0$ is
imaginary. This case arises for $k_x\neq k_y$ and leads to the same canonical
form (\ref{sim1}). Replacing $\omega_c^+\leftrightarrow\omega_c^-$ in Eqs.\
(\ref{tt1})--(\ref{tt2}) leads here to
\begin{equation}
h=\half({p}_+^2-{p}_-^2)-|\lambda|(q_+p_-+q_-p_+)\,,\label{hs2}
\end{equation}
where $p_\pm,q_\pm$ are hermitian. With a complex scaling $p_-\rightarrow
ip_-$, $q_-\rightarrow -iq_{-}$  Eq.\ (\ref{hs2}) becomes identical to Eq.
(\ref{hs1}) with an imaginary $\lambda$. The remaining equations remain then
unchanged but lead to exponentially increasing or decreasing evolutions for the
operators ${b}'_\pm$, $\bar{b}'_\pm$, corresponding to a ``boost'' applied to a
free motion.

L) This is an exceptional critical point where cases B,C,D,E,h,i,j and k
merge. Here $k_y=0$, $k_x=-4\omega^2$ (i.e., $k'_y=\omega^2$,
$k'_x=-3\omega^2$) or vice-versa, implying $|\omega|=\omega_c^+=\omega_c^-$ and
$\lambda_\pm=0$: All four eigenvalues of $\tilde{\cal H}_c$ vanish.
Nonetheless, $\tilde{\cal H}_c$ is still of rank $3$, implying that it becomes
similar to a full Jordan block ($d_\nu=4$):
\begin{equation}
\tilde{\cal H}_c\sim
\left(\begin{array}{cccc}0&1&0&0\\0&0&1&0\\0&0&0&1\\0&0&0&0\end{array}\right)
 \label{sim2}\end{equation}
instead of two free particle blocks, as would appear from Eq.\ (\ref{sim1}) for
$\lambda\rightarrow 0$. In this case the equations of motion cannot
be even partially decoupled. The transformation
\begin{eqnarray}
p_\pm&=&\delta_{\pm}(p_{x,y}+\omega q_{y,x}),\;\;
q_\pm={\textstyle\frac{1}{4\delta_\pm}}(3q_{x,y}-\omega^{-1}p_{y,x})
 \end{eqnarray}
with $\delta_+=1$, $\delta_-=2$,  allows to rewrite $h$ at this point as
\begin{equation}h=\half {p}_+^2-\omega q_+p_-\,,\label{hs3}\end{equation}
which is again a basic non-separable form: The ensuing equations of motion,
$\frac{dp_-}{dt}=0$, $\frac{dp_+}{dt}=\omega p_-$,
 $\frac{dq_+}{dt}=p_+$, $\frac{dq_-}{dt}=-\omega q_+$,
exhibit the structure of the Jordan form (\ref{sim2}) for $\omega\neq 0$ and
lead to a {\it polynomial evolution of third degree in $t$ for $q_-$:}
\begin{equation}\begin{array}{rcl}
p_-(t)&=&p_-,\;\;\;p_+(t)=p_++\omega t p_-,\\
q_+(t)&=&q_++p_+t+\half \omega t^2 p_-,\\
q_-(t)&=&q_--\omega tq_+-\half\omega t^2p_+
-{\textstyle\frac{1}{6}}\omega^2t^3p_-\end{array}\,.
\end{equation}
Coordinates $q_+$, $q_-$ experience then a {\it constant and linearly
increasing acceleration} respectively. In terms of the operators
$\bar{b}'_-=-\omega p_-$, $b'_+=ip_+$,
 $\bar{b}'_+=q_+$, $b'_-=iq_-/\omega$,
which satisfy $[b'_\nu,\bar{b}'_\mu]=\delta_{\mu\nu}$,
$[b'_\mu,b_\nu]=[\bar{b'}_\mu,\bar{b'}_\nu]=0$, we may also express Eq.\
(\ref{hs3}) as
 \begin{equation} 
 h=\bar{b}'_+\bar{b}'_--\half{b'}_+^2\,.\label{hs33}
 \end{equation}
Their equations of motion, $d\bar{b}'_-/dt=0$, $idb'_+/dt=\bar{b}'_-$,
$id\bar{b'}_+/dt=b'_+$, $idb'_-/dt=\bar{b}'_+$, follow {\it exactly} the Jordan 
form (\ref{sim2}). Note that the disappearance of one of the kinetic terms in
(\ref{hs3}) is not exceptional for a non-separable form: Eq.\ (\ref{hs1}) (and
hence (\ref{hs2})) can also be rewritten as ${p}_+^2-\lambda l$ if
$q_\pm\rightarrow q_\pm+\frac{1}{2\lambda}p_{\mp}$. We may also rewrite
(\ref{hs3}) with two kinetic terms with a similar transformation.

Eq.\ (\ref{sim2}) suggests that this case could be considered as a ``free
inseparable pair'', generalizing the free particle case where $\tilde{\cal
H}_c\sim(^{0\;1}_{0\;0})$. For a free particle $\tilde{\cal H}_c^2=0$ while
here $\tilde{\cal H}_c^4=0$.

\section{Conclusions}
We have first analyzed, within the formalism of ref.\ \cite{RK.05}, the
dynamics in general unstable quadratic bosonic forms, discussing the treatment
of the general non-diagonalizable case and determining the conditions for
dynamical stability and separability. We have then applied the formalism to the
basic problem of a particle in a general rotating quadratic potential, relevant
in the context of fast rotating condensates in harmonic traps and formally
equivalent to that of a charged particle in a uniform magnetic field in a
quadratic potential. The present analysis unveils the rich variety of behaviors
that can be exhibited by the unstable system, summarized in fig.\ \ref{f2},
together with some quite remarkable features, which could lead to observable
effects in fast rotating condensates. In particular, we have determined: a) The
regions of dynamical stability. Intrinsic motion in a rotating stable potential
remains dynamically stable at high frequencies, becoming unstable just in a
{\it finite} frequency window in the anisotropic case, whereas in a rotating
saddle potential it can become {\it dynamically stable} in a certain window
(Eqs.\ (\ref{uns}), (\ref{w4}) and fig.\ \ref{f1}); b) The regions in parameter
space where $H_{xy}$ can written as a sum of two independent quadratic systems
(separability), employing non-hermitian normal coordinates and momenta if
necessary (sector E), and those where such a representation {\it is not
feasible} (non-separability); c) The explicit transformations and final forms
for all cases, including the energy spectrum in the dynamically stable cases
and the ``minimally coupled'' standard forms and equations of motion in the
non-separable cases; d) The existence of an exceptional non-separable zero mode
case (point L in fig.\ \ref{f2}) where all eigenfrequencies vanish and the
Jordan Block has dimension 4. It is not equivalent to a standard zero frequency
mode and leads to coordinates evolving as a third degree polynomial in time.
These results indicate that similar peculiar effects can arise in more complex
unstable quadratic systems, which can be analyzed with the same general
formalism and techniques of sec.\ II.

RR and AMK are supported by CIC of Argentina.


\begin{thebibliography}{99}
\bibitem{RS.80} P.~Ring and P.~Schuck, {\it The Nuclear Many-Body Problem},
(Springer, NY, 1980).
\bibitem{BR.86} J.P. Blaizot and G. Ripka, {\it Quantum Theory of Finite
Systems} (MIT Press, MA, 1986).
\bibitem{GM.97} E.V.\ Goldstein, P.\ Meystre, Phys.\ Rev.\ A {\bf 55}, 2935
(1997).
\bibitem{PB.97} H.~Pu, N.P.~Bigelow, Phys.\ Rev.\ Lett.\ {\bf 80}, 1134
(1998);
C.K.~Law, H.\ Pu, N.P.~Bigelow, J.H.~Eberly, Phys.\ Rev.\ Lett.\ {\bf 79},
3105 (1997).
\bibitem{AK.02} S.~Alexandrov, V.V.~Kavanov, J.Phys.\ Condens.\ Matter 14,
L327 (2002); V.I.~Yukalov and E.P.~Yukalova, Laser Phys.\ Lett. 1, 50 (2004);
\bibitem{F.07}E.\ Fukuyama, M.\ Mine, M.\ Okumura, T.\ Sunaga, Y.\ Yamanaka,
Phys.\ Rev.\ A {\bf 76}, 043608 (2007);
M.\ Mine et al, Ann. Phys. {\bf 322}, 2327 (2007).
\bibitem{E.07}T.\ Sunaga et al, J.\ Low Temp Phys.\ 148, 381 (2007);
M.\ Mine et al, J.\ Low Temp Phys.\ 148, 331 (2007).
\bibitem{F.08}
Y.\ Nakamura, M.\ Mine, M.\ Okumura, Y.\ Yamanaka,
Phys.\ Rev.\ A {\bf 77}, 043601 (2008).
\bibitem{Sa.91} P.~Meystre, M.~Sargent, Elements of Quantum
Optics (Springer, NY, 1991).
\bibitem{GC.02} V.~Gurarie, J.T.~Chalker, Phys.\ Rev.\ Lett.\ 89, 136801
(2002).
\bibitem{P.97} I.A.\ Pedrosa, Phys.\ Rev.\ {\bf A55}, 3219 (1997).
\bibitem{Do.00} V.V. Dodonov, J. Phys. A {\bf 33}, 7721 (2000).
\bibitem{Ko.00} A.M.\ Kowalski et al, Phys. Lett. A {\bf 297}, 162 (2002).
\bibitem{RK.05} R.\ Rossignoli, A.M.\ Kowalski, Phys.\ Rev.\ {\bf A 72},
 032101 (2005).
\bibitem{D.05} A.\ Aftalion, X.\ Blanc, J.\ Dalibard,	
Phys.\ Rev.\ {\bf A} 71, 023611 (2005); S. Stock et al, Laser Phys.\ Lett. 2,
275 (2005).
\bibitem{BDS.08}
I.\ Bloch, J.\ Dalibard,   W.\ Zwerger, Rev.\ Mod.\ Phys.\ 80, 885 (2008).
\bibitem{Ft.01}
M.\ Linn, M.\ Niemeyer, A.\ L.\ Fetter, Phys.\ Rev.\ A {\bf 64}, 023602 (2001). 
\bibitem{Ft.07} A.\ L.\ Fetter, Phys.\ Rev.\ A {\bf 75}, 013620 (2007). 
\bibitem{O.04} M.\ \"O.\ Oktel, Phys.\ Rev.\ A {\bf 69}, 023618 (2004).
\bibitem{A.09}A.\ Aftalion, X.\ Blanc, N.\ Lerner,
Phys.\ Rev.\ A {\bf 79} 011603(R) (2009).
\bibitem{A.97} H. Attias, Y. Alhassid, Nucl.\ Phys.\ A {\bf 625}, 565 (1997).
\bibitem{RC.97} R.\ Rossignoli,  N.\ Canosa, Phys.\ Lett.\ B {\bf 394}, 242
(1997);
 R. Rossignoli, N.\ Canosa, P.\ Ring,
 Phys.\ Rev.\ Lett.\ {\bf 80} 1853 (1998); Phys.\ Rev.\ B {\bf 67},
 144517 (2003).
\bibitem{FK.70} A.\ Feldman and A.H.\ Kahn, Phys.\ Rev.\ B {\bf
1}, 4584 (1970).
\bibitem{V.56} J.G. Valatin, Proc.\ R.\ Soc.\ London, Ser.\ A {\bf 238}, 132 (1956).
\bibitem{L.30} L.\ D.\ Landau, Z. Physik {\bf 64}, 629 (1930).

\end{thebibliography}
\end{document}